# AN AUTOMATIC METHOD FOR DETERMINATION OF Lg ARRIVAL TIMES USING WAVELET TRANSFORMS

TIBULEAC, I. M. and HERRIN, E. T., Department of Geological Sciences, Southern Methodist University, Dallas, Texas, TX 75275-0395, ileana@passion.isem.smu.edu; herrin@passion.isem.smu.edu

## ABSTRACT

The regional phase Lg is often used to estimate location and magnitude for sources closer than 1500 km. The complexity of Lg waveforms makes it difficult to consistently determine a unique Lg arrival time, thus affecting source location with a single station or array. This study tests an automatic method for timing Lg arrivals using wavelet transforms to decompose the Lg signal into its components localized both in time and scale.

A Continuous Wavelet Transform (CWT) using a Daubechies order two (db2) wavelet is applied to 10 seconds of raw data containing the start of Lg. Initial positioning of the window is obtained using standard Lg travel time tables. The coefficients at scale 16 from the db2 decomposition are squared and the resulting time series is represented by an approximation of the 4'th level Discrete Wavelet Transform (DWT) using a Haar wavelet. A threshold detector is then applied to the resulting time series to determine the Lg arrival time. The method was tested using well located earthquakes and explosions from known mines ( $m_b < 4.0$ ), recorded on the vertical components at TXAR (Lajitas, Texas) and PDAR (Pinedale, Wyoming) arrays. The Lg arrival time was automatically picked with a standard deviation of less than 1.5 seconds (less than 10 km

2location error) for well known locations. Location errors are larger with the increase of distance and smaller with the increase in signal to noise ratio of events.

## INTRODUCTION

Lg is one of the most used regional phases, both in location and discrimination, and dominant for many events up to 1500 km epicentral distance. Lg is made up of short period guided waves including scattering contributions from crustal heterogeneity, modeled as a sequence of multiply reflected postcritical shear waves trapped in the crustal wave-guide (Xie and Lay, 1994), or as a superposition of higher mode surface waves (Herrmann and Kijko, 1983; Bollinger, 1979). The properties of the phase include a period of 1 to 7 seconds, a tendency to horizontal polarization, and a mean surface velocity of about 3.5 km/s (Herrin and Minton, 1960). The physical mechanism of Lg generation is still under study (Gupta *et al.*,1997; Patton and Taylor, 1995; Zhang and Lay, 1995; Xie and Lay, 1994) and there is no widely accepted automatic way of picking its arrival.

Different types of seismic phase detectors, in time-domain and frequency-domain and using particle motion processing or pattern matching were previously developed (see Withers *et al.*, 1998, for a good review). None of them have proved to be acceptable for all situations. In their paper, containing the study of an automated, near-real-time, waveform correlation event-detection and location system, Withers *et al.*, (1998), use the idea of detecting transients in the frequency content of seismograms for one of the digital algorithms which they test in order to detect seismic phases for teleseismic events. Their solution was to use adaptive STA/LTA processing (variable length window function of the dominant frequency of the signal) and they were able to enhance a large variety of phase arrivals over a wide range of source, path, receiver and noise scenarios. However, the length of the windows used affected the precision of their estimates, even in the case of adaptive processing. Envelope processing, suitable for cases when original waveforms are very complex, was used by Husebye *et al.*, (1998), in local event semi-automatic locations, for a frequency band between 3 and 6 Hz. Their procedure, while simple and suitable for real-time monitoring purposes, was also affected by the lack of resolution due to windowing. Wavelet transform methods were used for P and S phase identification in three - component seismograms by Anant and Dowla (1997). Assuming that strong features of the seismic signal appear in the wavelet coefficients across several scales, they used a multi-scalar representation to construct "locator" functions that identify the P and S arrivals and developed the first detector based on an alternative method of signal processing using wavelet decomposition of the signal.

We present an automatic method of picking Lg phases based on detecting transients in frequency content of seismograms and obtain the envelopes of the waveforms with a discrete wavelet decomposition method. The wavelet analysis is a windowing technique, with variable sized regions, that allows the use of long time intervals in instances for which we want precise low frequency information and shorter intervals where we want high frequency information. The method is based on the capacity of the wavelet transform for detecting changes in the nature of a process and to express frequency (scale) content changing in time, which is not possible with Fourier analysis or even windowed Fourier analysis (Misiti *et al.*, 1996; Strang and Nguyen, 1996). We make the assumption that Lg, being composed of surface waves following the P and S coda, constitutes a change in process and is likely to be seen as a change of pattern in the time - scale representation of the continuous wavelet transform (CWT). We also assume that the Lg duration on the seismogram is of the order of at least 3 seconds, based on the fact that the formation of Lg requires a significant amount of energy propagating as shear waves trapped in the crustal wave - guide (see also Xie and Lay, 1994). The automatic detection method was tested on events recorded at two arrays: TXAR (Lajitas, Texas) and PDAR (Pinedale, Wyoming). The location and station configuration of the arrays are presented in Figure 1.

## THE DATA

The data set of small magnitude events ( $m_{bLg} < 4.0$ ) consists of explosions from four mining districts and aftershocks of the April 14, 1995, $m_w$ = 5.8 Alpine (West Texas) earthquake, recorded at TXAR, and of a set of explosions from Black Thunder Mine (Powder River Basin, Wyoming), recorded at PDAR. The data from the vertical component of one of the elements TX08 or TX09 at TXAR and PD03 at PDAR were analyzed for each event. Locations of the earthquakes and of the mines, the mine status (multiple mines or single mine), the distance and backazimuth to the respective array, as well as the number of events used in this study are presented in Table 1. Locations relative to TXAR and PDAR of the events used to assess the Lg detector are presented in Figure 2. Red squares are mine locations, the blue square represents the Alpine aftershocks locations.

5The 16 Alpine aftershocks were located by USGS (Preliminary Determination of Epicenters Catalog) northeast of TXAR, at a distance of 108 km, backazimuth 19 degrees. They were used because they are well located relative to the main shock.

Two clusters of mines were found in Mexico south and southeast of TXAR by Sorrells *et al.* (1997). For the 21 events from Minas de Hercules iron mines, Mexico (approximately 150 km distance from TXAR, 185 degrees backazimuth) and 22 events from MICARE (Minera Carbonifera de Rio Escondito, Mexico), at approximately 329 km from TXAR, backazimuth 110 degrees, we did not have ground truth. We believe that these events originate from a number of mines for which we do not have precise locations.

Nine explosions from the copper mine at Tyrone, New Mexico (579 km distance from TXAR, 311 degrees backazimuth) were confirmed by the engineers at the mine as either simultaneous or delay fired blasts. From the Morenci copper mine, Southeastern Arizona, (681 km from TXAR, backazimuth 309 degrees), 8 confirmed delay fired explosions were analyzed.

The seven events from the Black Thunder mine, at 360 km distance from PDAR, backazimuth 73 degrees, were well documented blasts (Stump and Pearson, 1997) located in the same pit; one was a simultaneous explosion, the others were delay fired explosions.

Bonner et al, (1997), observed that a relatively smaller yield in a simultaneous explosion from Tyrone produced a similar Lg magnitude to a larger delay fired explosion. For this reason we did not use the magnitude to characterize the events, but the signal to noise ratio (SNR) instead. For this study, the SNR is defined as the ratio of the sum of the squares of the coefficients of the continuous wavelet transform (CWT) at scale 16 in a 5 seconds window after and immediately before the Lg arrival.

# THE METHOD

## Wavelets and Fourier Analysis

Fourier analysis assumes that the spectral content of the data does not change with time within the analysis window. Thus frequencies present at one time are presumed to be present with the same intensities at all times. In fact, the time varying spectrum is undefined in Fourier analysis. However, seismic data is clearly time-varying in frequency, a fact that has been realized for some time. Previous researchers have addressed this problem by computing Fourier spectra with overlapping windows and piecing them together. Such an approach is not altogether satisfactory, is costly in computation time and difficult to optimize, although it is headed in the right direction - the direction of wavelet analysis. In wavelet analysis, unlike the Fourier transform which is global in frequency and independent of time, the wavelet transform is a localized transform in space (time) and scale (generalized frequency). Figure 3 is an example of both transforms applied to a signal containing the same frequencies superposed as in Figure 3 a, or successive as in Figure 3 b. The sampling rate is 40 samples/second and a Hanning window is applied before calculation of the Fast Fourier Transform presented in subplots 2a and 2b. The Daubechies order 10 (db10) continuous wavelet transform ( see Misiti et al., 1996; Strang and Nguyen, 1996), in subplots 3a and respectively 3b, shows the frequency content changing in time, which is not possible to identify in the Fourier analysis. The $y$ axis in subplots 3a and 3b represents the scale; low scales correspond to high frequencies, larger scales correspond to low frequencies.

A first approach to the problem of obtaining the frequency content of a process locally in time was a windowing technique applied to the signal, called the Short Time Fourier Transform (STFT) that maps a signal into a two-dimensional function of time and frequency (Chui, 1997; Kumar and Foufoula-Georgiou, 1996). However, the time/frequency window of the STFT is fixed and hence unsuitable for detecting signals with both high and/or low frequencies.

In essence, wavelet analysis is an extension of the STFT in such a way that the time/frequency window is no longer fixed, but can be varied. The use of long time intervals where we want more precise low frequency information and shorter intervals where we want high frequency information is accomplished through the introduction of a location parameter and a scale parameter as opposed to the Fourier method which has only a scale parameter. One of the two parameters is the translation or location parameter as in the windowed Fourier transform case. The other parameter is a dilation or scale parameter, the latter corresponding to the frequency $\omega$ in the Fourier case.

A major advantage of analyzing a signal with wavelets as the analyzing kernels is that it allows for the study of signal features locally with a detail matched to their scale, i.e. broad features at a large scale and fine features on small scales. This property is especially useful for signals that are non-stationary, have short-lived transient components, have features at different scales or have singularities, discontinuities in higher derivatives and self-similarity (Kumar and Foufoula-Georgiou, 1996).

8## Basic wavelets theory

The essence of wavelet analysis is to expand a given function f(t) as a sum of 'elementary' functions called wavelets. Wavelets are functions that satisfy certain requirements. The name wavelet comes from the two requirements that they should integrate to zero, waving above and below the x-axis and that the function has to be well localized. A wavelet is a waveform of effectively limited duration that has an average value of zero. They tend to be irregular and asymmetric (see Strang and Nguyen, 1996).

The wavelets are themselves derived from a single function $\Psi$, called the mother wavelet, by translations and dilations (Priestley, 1996). The mother wavelet is chosen so that it satisfies the following conditions:

$$\int_{-\infty}^{\infty} \psi(t)dt = 0$$

$$\int_{-\infty}^{\infty} |\psi(t)|dt < \infty$$

$$\int_{-\infty}^{\infty} \frac{|\Psi(\omega)|^2}{|\omega|} d\omega < \infty$$

where $\Psi(\omega)$ is the Fourier transform of $\psi(t)$ and $\Psi(0) = 0$. The mother wavelet is 'well localized' and decays 'quickly' to zero at a suitable rate (has compact support).

A classical mother wavelet is the Haar function defined by:

$$\psi^H(t) = \begin{cases} 1, & 0 \leq t \leq 1/2 \\ -1, & 1/2 \leq t \leq 1 \\ 0, & otherwise \end{cases}$$

and shown in Figure 4, subplot 1. The Haar wavelet is not continuous, therefore the choice of the Haar basis for representing smooth functions, for example, is not natural and economic. However, when looking for a sudden change in the signal, the Haar wavelet is recommended (Misiti *et al.*, 1996).

Given a mother wavelet, a doubly infinite sequence of wavelets can be constructed by applying varying degrees of translations and dilations to the mother wavelet. For real *a, b* ($a \neq 0$)

$$\psi_{a,b}(t) = |a|^{-1/2} \psi(\frac{t-b}{a})$$

so that the parameter *a*, usually restricted to positive values, represents the scale parameter and *b* the translation parameter. The normalizing factor $|a|^{-1/2}$ is included so that

$$\int |\psi_{a,b}(t)|^2 dt = \int |\psi(t)|^2 dt .$$

This equality is always true for $a > 0$.

Figure 4, subplot 2 presents the compactly-supported wavelet db2 of Ingrid Daubechies (named Daubechies order two wavelet). The names of the Daubechies wavelets ( see Misiti et. al, 1996) are written *dbN*, where *N* is the order and *db* the ìsurnameî of the wavelet.

Selection of the wavelet shape is an important decision. Each may show a part of the reality with specificity, and each may reveal something that the others had concealed. Choosing the wavelet is an advantage of the wavelet application and it allows for flexibility to adjust the analysis depending on the signal type, but there is no generally accepted procedure for doing so. Anant and Dowla (1997), for P and S arrivals, matched the wavelets with the arrival shape to help locate the arrival in the seismic signal. They also used a pseudo-wavelet constructed from the P arrival shape to locate the S arrival. The Lg arrival is too complicated for this procedure, so that we used a very simple wavelet, db2, composed of only one cycle, for our study. Also, for the best characterization of the beginning of the Lg arrival, the Haar wavelet (db1) was considered the most suitable.

### . **Wavelet analysis**

As Fourier analysis consists of breaking up a signal into sine waves of various frequencies, wavelet analysis is the breaking up of a signal into shifted (translated) and scaled (dilated or compressed) versions of the mother wavelet, each multiplied by an appropriate coefficient. Any function $f(t) \in L_2$, the space of square integrable functions, can be expressed as a linear superposition of $\{\psi_{a,b}(t)\}$.

The process of Fourier analysis is represented by the Fourier transform:

$$F(\omega) = \int_{-\infty}^{\infty} f(t)e^{-i\omega t}dt$$

which is the sum over all time of the signal f(t) multiplied by a complex exponential. The Fourier coefficients F($\omega$), when multiplied by a sinusoid of appropriate frequency $\omega$, yield the constituent sinusoidal components of the original signal.

Similarly, the Continuous Wavelet Transform (CWT) is defined as the sum over time of the signal multiplied by scaled, shifted versions of the wavelet function $\psi$ :

$$C(scale, position) = \int_{-\infty}^{\infty} f(t)\psi_{a,b}(t)dt$$

where C are the wavelet coefficients, functions of scale and position, which multiplied by the appropriated scaled and shifted wavelet yields the constituent wavelets of the original signal ( Misiti *et al.*, 1996). The continuous wavelet transform can operate at every scale and the analyzing wavelet is shifted smoothly over the full domain of the analyzing function. This distinguishes it from the Discrete Wavelet Transform (DWT). A good review of theoretical aspects related to DWT was presented by Anant and Dowla, (1997) and Priestley, (1996).

The advantage of the DWT is that, using dyadic scales and positions (based on powers of two) the analysis is more efficient and not redundant. For fixed values $a_0$ and $b_0$, if $a = a_0^m$, $b = nb_0 a_0^m$, $n,m = 0, \pm 1, \pm 2, ...$, and for $a_0 = 1/2$ and $b_0 = 1$, the Daubechies discrete wavelet family is:

$$\psi_{m,n}(t) = 2^{m/2} \psi(2^m t - n)$$

This class of wavelets forms a complete orthonormal basis for the space of square integrable functions, so f(t) admits the representation:

$$f(t) = \sum_{m=-\infty}^{\infty} \sum_{n=-\infty}^{\infty} c_{m,n} \psi_{m,n}(t) \quad \text{with}$$

$$c_{m,n} = \int_{-\infty}^{\infty} f(t) \psi_{m,n}(t) dt.$$

The mother wavelet $\psi(t)$ is dilated by the factor $2^{-m}$ and translated to the position $n \cdot 2^{-m}$. $\psi_{m,n}(t)$ is "localized" around the time point $t = n \cdot 2^{-m}$ and hence the wavelet coefficient $c_{m,n}$ depends only on the local properties of f(t) in the neighborhood of $t = n \cdot 2^{-m}$. A very practical filtering algorithm developed in 1988 by Mallat (see Misiti *et al.*, 1996; Strang and Nguyen, 1996) yields a fast wavelet transform. Two filters, high pass and low pass, are used to obtain respectively the detail (D) and the approximation (A) of the signal. Downsampling ( throwing away every second data point) is introduced to keep the total number of data points constant. The decomposition process can be iterated, with successive approximations (A) being decomposed in turn, so that the signal is broken down into many lower resolution components (the wavelet decomposition tree). A disadvantage of DWT, from a seismological point of view, is the downsampling steps that result in the length of detail j+1, j≥1, j ∈ N being half the length of detail j (see also Anant and Dowla, 1997). As an example, for a signal with a sampling rate of 40 samples/second, detail 1 and approximation 1 will have 20 samples/sec, and detail 4 and approximation 4 will have 2.5 samples /second. When looking for the overall trend of the signal, equivalent to its behavior at low frequencies, the trend is more clear with each approximation, when j increases, and there is a trade-off between the resolution (sampling rate) and the smoothness of the approximation.

Scaling a wavelet means stretching or compressing it. In wavelet analysis the scale factor is also related to the frequency of the signal. Higher scales correspond to the most ìstretchedî wavelets and the coefficients measure coarse signal features. Low scales correspond to compressed wavelets which analyze rapidly changing details associated with high frequencies. The fact that the wavelet basis functions are indexed by two parameters (*a, b or m, n*) whereas the Fourier basis functions are indexed by the single parameter *ω*, means that wavelet transforms (or coefficients) are characteristics of the local behavior of the function whereas the Fourier transforms (or coefficients) are characteristics of the global behavior of the function.

The concept of "frequency" is related purely to the complex exponential function and has no precise meaning when applied to other families of functions, unless a more general concept of "frequency" can be defined (Priestley, 1996). In a loose sense, the wavelet parameters *a, b, or m, n* may be identified with generalized frequency and with time: when parameter *a* decreases, the "oscillations" become compressed in the time domain, i. e., they exhibit "high frequency" behavior, and as the parameter *a* increases, the "oscillations" exhibit ìlow frequencyî behavior.

**The method**

A plot of typical raw waveforms from all locations is presented in Figure 5. The first five (5) subplots represent events recorded at TXAR. Subplot 6 represents an event recorded at PDAR. The search window of 10 seconds (400 samples) is also represented and the amplitude in nanometres @ 1 second is counts x 0.0073 for TXAR and counts x 0.005 for PDAR.

A continuous wavelet transform is applied to the raw data, for a scale from 1 to 20, with step 1, using the db2 wavelet. We used the db2 CWT in order to maintain constant sample rate of the signal. The method is illustrated in Figures 6-9 which show the processing steps for two Tyrone events recorded at the station TX08. Figure 6 shows the October 11, 1996 simultaneous explosion and Figure 8 shows the September 9, 1997 delay fired explosion. In Figures 6 and 8, subplot 1 contains the raw data, with the mean removed, for each of the two explosions at Tyrone. The *x* axis represents samples (40 samples/sec) in all subplots. Absolute values of the CWT coefficients using db2 wavelet are presented in subplot 2 of each figure. The *y* axis represents scale. The color of each point represents the magnitude of the wavelet coefficient. Light colors correspond to large

14coefficients, dark colors correspond to small coefficients. Low scales are indicative of high frequency, high scales are indicative of low frequency. A clear change in color marks the Lg arrival.

The coefficients at scale 16 were empirically chosen as the best to be analyzed for all the events. The range of frequency they correspond to is 0.9 to 2.1 Hz, centered on 1.5 Hz (see Chui, 1997), which is a band with significant energy in the Lg Fourier spectrum for all the events in this study as well as for events from other regions. As an example, the corner frequency for events with magnitude less than 4.0 in Eastern Canada, for distances between 100 and 900 km, was calculated by Hasegawa, (1983) to be higher than 2 Hz. Also, 1 sec period Lg waves were used by Nuttli (1986) for estimating seismic magnitudes because their anelastic attenuation is small in shield and other geologically stable regions. Studies of the decay of spectral energy of Lg waves with distance for different frequencies and different regions generally show a less rapid decay in the range of 0.5 Hz to 2 Hz (Campillo *et al.*, 1985; Herrmann *et al.*, 1997). Subplot 3 of both Figures 6 and 8 presents the coefficients at scale 16 ( the sampling rate of the initial signal is preserved).

Figures 7 and 9 represent the continuation of the analysis of the Tyrone events. Squared coefficients at scale 16 are presented in subplot 1 of each figure. The 10 sec search window for Lg is expanded in subplot 2 for each event. A discrete Haar wavelet decomposition of the signal in subplot 1 is performed and the approximation at level 4 in the 10 seconds search window is represented in subplot 3 in Figures 7 and 9. The time series are normalized to the maximum value of Lg amplitude over a 15 second window, starting at the same point as the 10 second search window. The level 4, selected empirically, using a trial and error procedure, has a sampling rate of 2.5 samples/second and an initial error in picking the Lg arrival of 0.4 seconds representing an error in distance estimated from Lg minus P times of approximately 2.5 to 2.8 km.

An empirically chosen 0.25 threshold detector is applied in determining the Lg arrival time for all the events. The threshold was chosen as the one that gives lowest Lg minus P travel time standard deviations for all locations. The 0.25 threshold exceeds the mean plus one standard deviation of the detail 5, (the ìnoiseî for the approximation at level 4), normalized to the maximum of approximation at level 4 for the majority of the events.

The detector includes a checking algorithm designed to verify that the amplitude of the Lg envelope exceeds the threshold for several seconds. For each envelope value that exceeds the threshold, a linear interpolation is performed in a 4 seconds window centered on the respective time, and if at least 70% of the values exceed the threshold, the point is considered to be the time of Lg arrival. This procedure is designed to eliminate the possibility of picking S arrivals coming before Lg and is based on the assumption that the Lg duration is at least 3 seconds.

## RESULTS AND DISCUSSION

The standard deviation of the Lg and P arrival time difference, expressed in km, for each location as a function of the distance to the respective array is presented in Figure 10. Table 1 presents the numerical values of the standard deviation (km) for each location. The standard deviation was calculated with the formulas:

$\sigma_d = 7.58 * \sigma_{\Delta t_{Lg-P}}$ (km) for Pg first arrival, i.e. for the Alpine aftershocks and for the Minas de Hercules events and

$\sigma_d = 6.22 * (6.5 + \sigma_{\Delta t_{Lg-P}})$ (km) for Pn first arrival for all the other events, where $\Delta t_{Lg-P}$ is the travel time difference between the Lg and P arrivals. The Lg arrival time was automatically picked with a standard deviation of less than 10 km for all the single mines. The Mexico mines, considered to be multiple mines, have standard deviations greater than 10 km and less than 15 km.

The plot of the standard deviation as a function of Lg SNR threshold, in Figure 11, for all the single mine locations generally shows smaller errors with increase in SNR. However for larger SNR fewer events were considered. There is no significant variation in the standard deviation with SNR for the Alpine aftershocks.

The raw data with the mean removed for the October 11, 1996 simultaneous explosion at Tyrone are presented in Figure 12, subplot 1. Subplot 2 shows the db2 CWT coefficients at scale 16 and the waveforms filtered between 0.9 and 2.1 Hz are presented in subplot 3. A zero phase, forward and reverse, 4'th order Butterworth filter was applied on the raw data in subplot 1 and, due to the very narrow frequency range, the resulting waveforms loose features. Also, the amplitude difference between the Lg signal and the ìnoiseî before Lg is much less than in the case of wavelet coefficients. We conclude that the wavelet transform is a more efficient way of filtering data in narrow frequency bands.

Comparing the shapes of the Lg envelopes for the simultaneously fired and for the delay fired explosion at Tyrone (Figures 7 and 9, subplots 3) it is seen that the Lg signal from the simultaneous explosion is impulsive, followed by an exponential decay, like a delta function convolved with an exponential, while the delay fired Lg is emergent, like a boxcar convolved with an exponential. The same difference in shape is observed for the approximations of Lg squared CWT coefficients for the events at Black Thunder Mine, not only at scale 16, but also for the raw data. We compare in Figure 13 two Black Thunder events, a simultaneous explosion and a delay fired explosion. The raw data with the mean removed for the August 24, 1995 simultaneous explosion is presented in subplot 1 and the raw data with the mean removed of the delay fired explosion on June 23, 1995 is presented in subplot 3. The Haar approximation at level 4 of the squared CWT coefficients at scale 16, in subplots 2 and 4, is different for each type of explosion; i.e., a sharp increase in Lg for the simultaneous explosion and a less rapid rise of Lg followed by a slower decay of amplitude for the delay fired blast. If the length of the boxcar could be considered proportional to the duration of the source, then it is possible that this type of analysis could be used as a discriminant between simultaneous explosions and delay fired explosions, provided that both occur in the same location. It could also be an additional tool for discrimination between simultaneous and delay fired explosions besides the study of long period regional surface waves, described by Stump and Pearson (1997), applicable for the large cast shots in same data set.

## CONCLUSIONS

The automatic method for determination of Lg arrival times presented here results in a location error of about 10 km or less for events at epicentral distances from 360 km up to 700 km. Larger location errors at the Micare coal mines and the Minas de Hercules iron mines are probably caused by the existence of clusters of mines. A future study will focus on the statistical discrimination of individual mines in such a cluster.

The method described here allows extraction of meaningful signal attributes from single station vertical records using a set of conditions of which many are empirically established. The use of different wavelets and, specifically, different scales will be investigated in future studies. An automatic method to choose the threshold for the threshold detector as well as a better way of recognizing Lg than linear interpolation are under study.

**ACKNOWLEDGMENTS**

We thank Dr. Henry Gray, Dr. Stephen Wiechecky, Dr. Jessie Bonner, Dr. Mike Sorrels and Paul Golden for discussions on the manuscript, Dr. Brian Stump for the Black Thunder data. We also thank Dr. Gilbert Strang and Dr. Aaron Velasco for helpful suggestions and comments. The research was supported by the Defense Special Weapons Agency Grant 001-97-I-0024.

**FIGURE CAPTIONS**

**Figure 1**. The location and station configuration of the TXAR (Lajitas, Texas) and PDAR (Pinedale, Wyoming) arrays. Triangles represent three components stations (3-C), squares represent one component (1-C) stations

**Figure 2**. Locations relative to TXAR and PDAR of the events used to assess the Lg detector. Red squares are mine locations, the blue square represents the Alpine aftershocks locations.

**Figure 3**. Example of Fourier transform and Continuous Wavelet Transform applied to a signal containing the same frequencies: superposed as in **subplot 1a** or successive as in **subplot 1b**. The *x* axis represents time in samples (sample rate = 40 samples/sec) in all subplots.
**Subplots 2a and 2b**: The frequency content of the signals obtained in each case by Fast Fourier Transform. The two frequencies in the signals are clearly observed in both cases.
**Subplots 3a and 3b**. The Daubechies order 10 (db10) Continuous Wavelet Transform applied to both signals shows the frequency content of the signal in time. Subplot 3a shows the two frequencies superposed, in subplot 3b, the arrival time of the second frequency can be seen. The *y* axis represents the scale, low scales correspond to high frequencies, larger scales correspond to low frequencies

**Figure 4**. **Subplot 1**. A classical mother wavelet: the Haar function.

**Subplot 2.** The compactly-supported wavelet db2 (named Daubechies order two wavelet).

**Figure 5**. Typical raw waveforms from all locations. . The *x* axis represents samples (40 samples/sec) for all subplots. On the *y* axis, the amplitude in nanometres @ 1 second is counts x 0.0073 for TXAR and counts x 0.005 for PDAR. The search window of 10 seconds (400 samples) is also represented together with the value of the SNR for Lg at scale 16.

**Subplot 1 - 5**. Events recorded at TXAR.

**Subplot 6.** An event recorded at PDAR

**Figure 6**. The October 11, 1996 Tyrone simultaneous explosion.

**Subplot 1**. Contains the raw data, with the mean removed, recorded at the station TX08. The *y* axis represents amplitude in nanometres @ 1 second is counts x 0.0073. The *x* axis represents samples (40 samples/sec) for all subplots.

**Subplot 2**     A continuous wavelet transform is applied to the raw data, for a scale from 1 to 20, with step 1, using the db2 wavelet. Absolute values of the CWT coefficients are presented and the *y* axis represents scale. The color of each point represents the magnitude of the wavelet coefficient. Light colors correspond to large coefficients, dark colors correspond to small coefficients. Low scales are indicative of high frequency, high scales are indicative of low frequency. A clear change in color marks the Lg arrival.

**Subplot 3**. The coefficients at scale 16 ( the sampling rate of the initial signal is preserved).

**Figure 7**. The next steps of the analysis in Figure 6.

**Subplot 1**. Squared coefficients at scale 16. The *x* axis represents time in samples (sample rate = 40 samples/sec) in all subplots.

**Subplot 2**. Coefficients at scale 16 in the expanded 10 seconds search window for Lg.

**Subplot 3**. The approximation at level 4 of the squared coefficients obtained with a discrete Haar wavelet decomposition, represented in the search window. The time series are normalized to the maximum value of Lg amplitude over a 15 second window, starting at the same point as the 10 second search window

**Figure 8**. Same as in Figure 6, for the September 9, 1997 delay fired explosion.

**Figure 9**. Same as in Figure 7, for the September 9, 1997 delay fired explosion.

**Figure 10**. The standard deviation of the Lg and P arrival time difference, expressed in km, for each location as a function of the distance to the respective array.

**Figure 11**. Standard deviation as a function of Lg SNR threshold for all the single mine locations.

**Figure 12**. **Subplot 1**. The October 11, 1996 Tyrone simultaneous explosion: raw data with the mean removed. The *x* axis represents time in samples (sample rate = 40 samples/sec) in all subplots.

**Subplot 2**. db2 CWT coefficients at scale 16 of the signal in subplot 1.

**Subplot 3**. Same signal filtered between 0.9 and 2.1 Hz using a zero phase, forward and reverse, 4'th order Butterworth filter.

**Figure 13.** Compare the Lg arrivals of two Black Thunder events, a simultaneous explosion and a delay fired explosion.

**Subplot 1**. Lg raw data with the mean removed for the August 24, 1995 simultaneous explosion. The *x* axis represents samples (sample rate = 40 samples/sec) in all subplots.

**Subplot 2**. Haar approximation at level 4 of the Lg squared db2 CWT coefficients at scale 16, in the search window, for the signal in subplot 1.

**Subplot 3**. Lg raw data with the mean removed of the delay fired explosion on June 23, 1995.

**Subplot 4**. Haar approximation at level 4 of the Lg squared db2 CWT coefficients at scale 16, in the search window, for the signal in subplot 3.

|    | Location | Latitude | Longitude | Status | Array | Distance (km) | Back-azimuth (deg) | Number of events | Std. dev. automatic scale 16 (km) |
|----|----------|----------|-----------|--------|-------|---------------|--------------------|------------------|----------------------------------|
| 1. | Alpine, Texas | 30.3 | -103.3 | A | TXAR | 108 | 19 | 16 | 3.5 |
| 2. | Minas des Hercules (Mexico) | 28.1 | -103.8 | M | TXAR | 150 | 185 | 21 | 15 |
| 3. | Micare Coal Mines (Mexico) | 28.3 | -100.5 | M | TXAR | 329 | 110 | 22 | 13 |
| 4. | Tyrone | 32.7 | -108.4 | S | TXAR | 579 | 311 | 9 | 5 |
| 5. | Morenci | 33.1 | -109.4 | S | TXAR | 681 | 309 | 8 | 9 |
| 6. | Black Thunder | 43.7 | -105.3 | S | PDAR | 360 | 73 | 7 | 2.5 |

A - Aftershocks of the April 14, 1995, m$w$ = 5.8 Alpine (West Texas) earthquake
M - Multiple mines
S - Single mine

**Table 1**

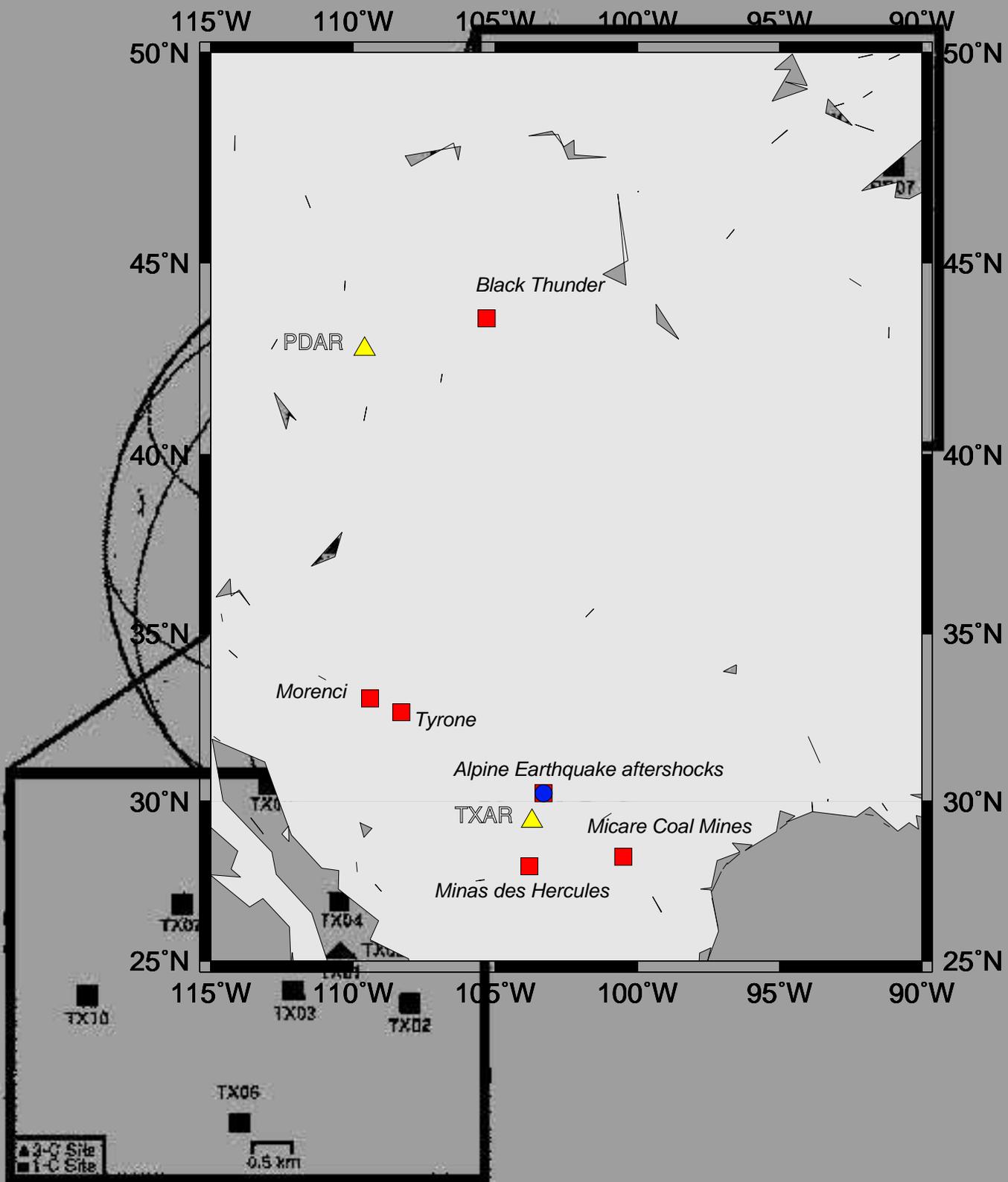

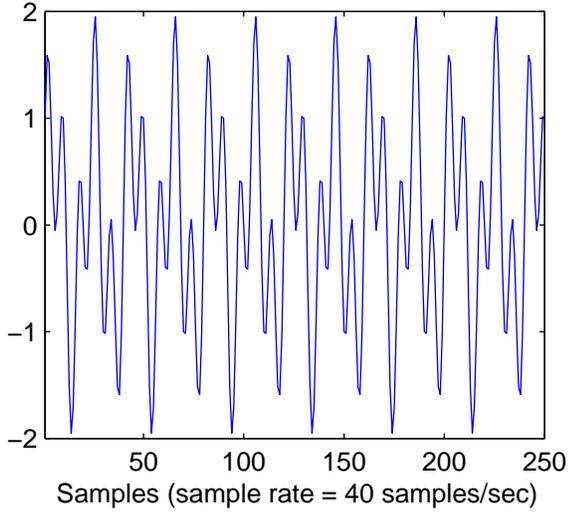
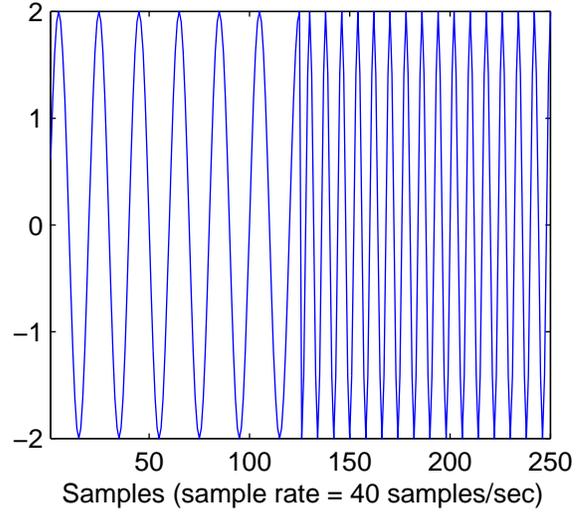
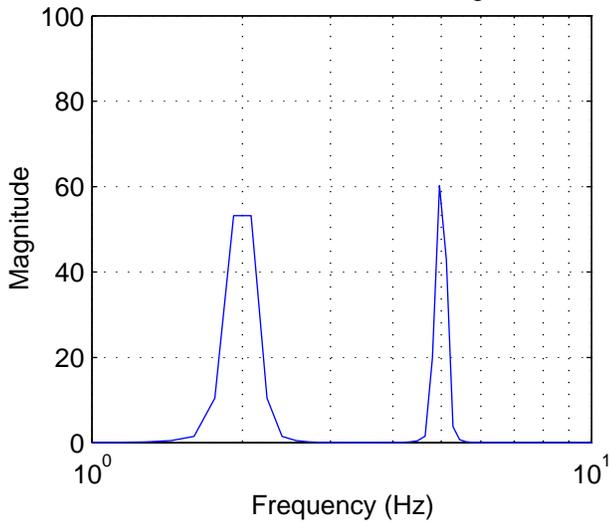
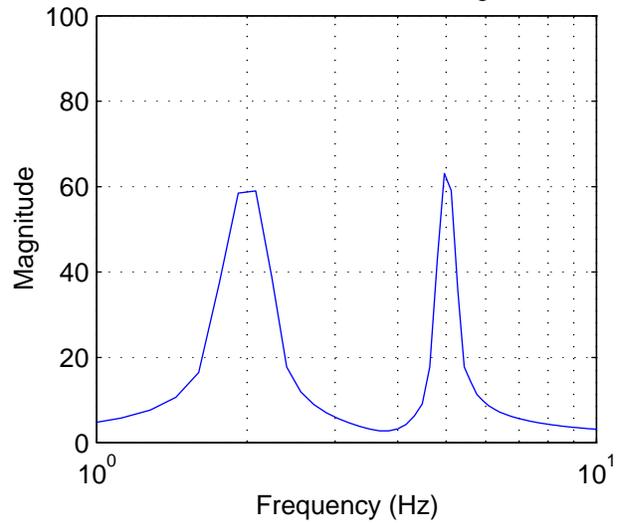
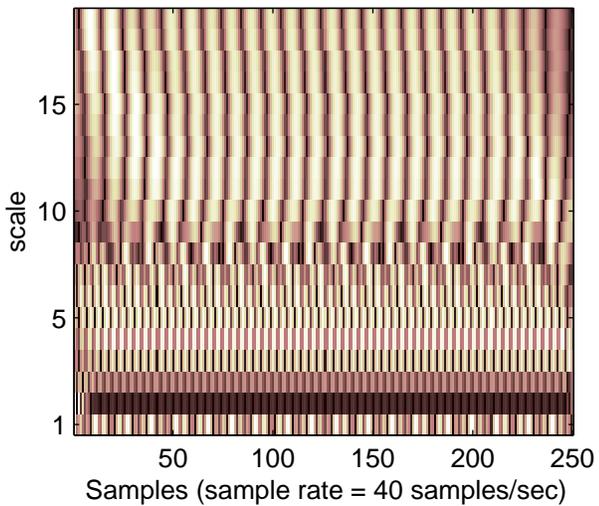
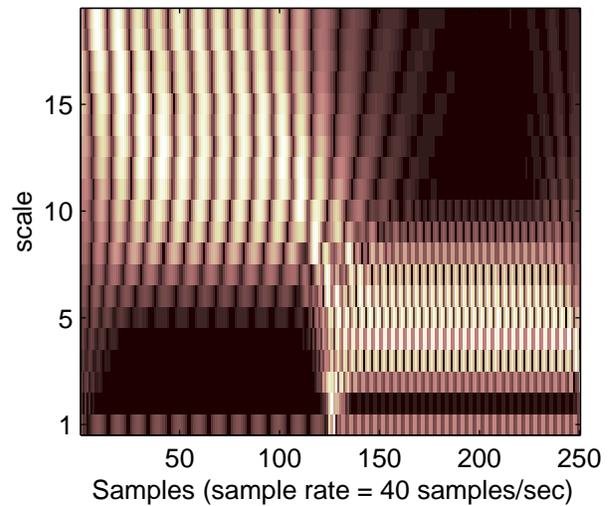

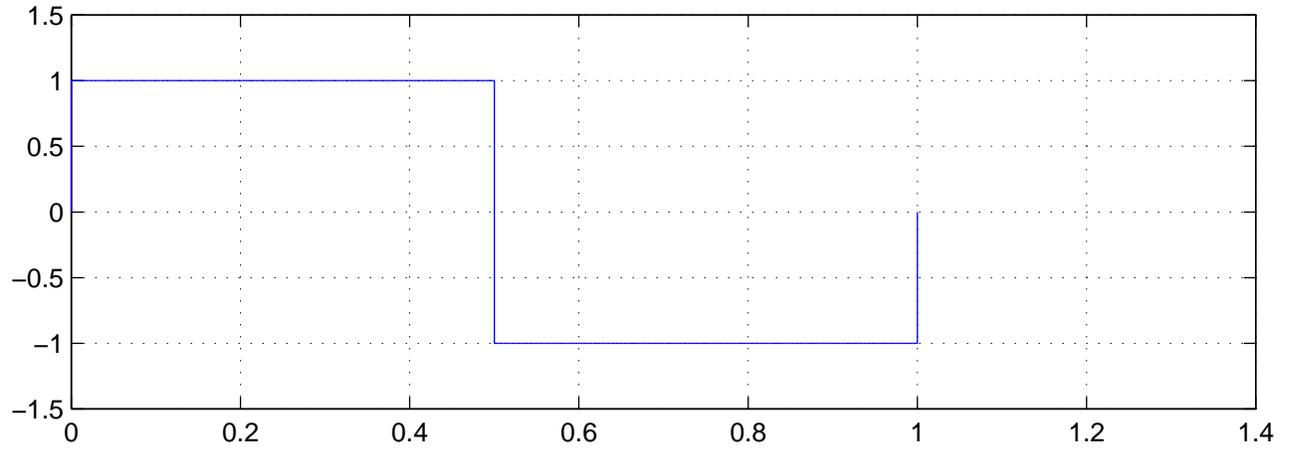
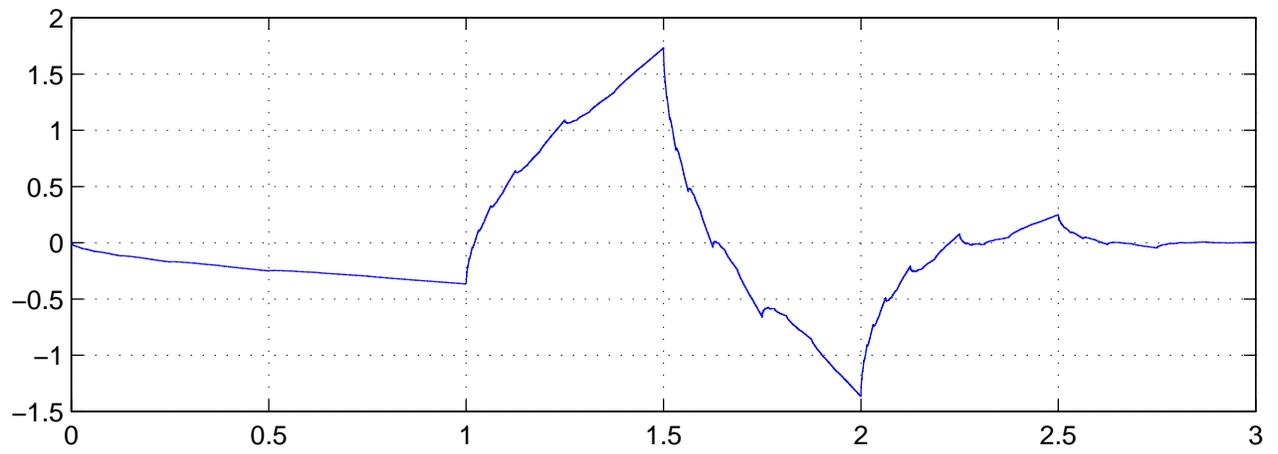

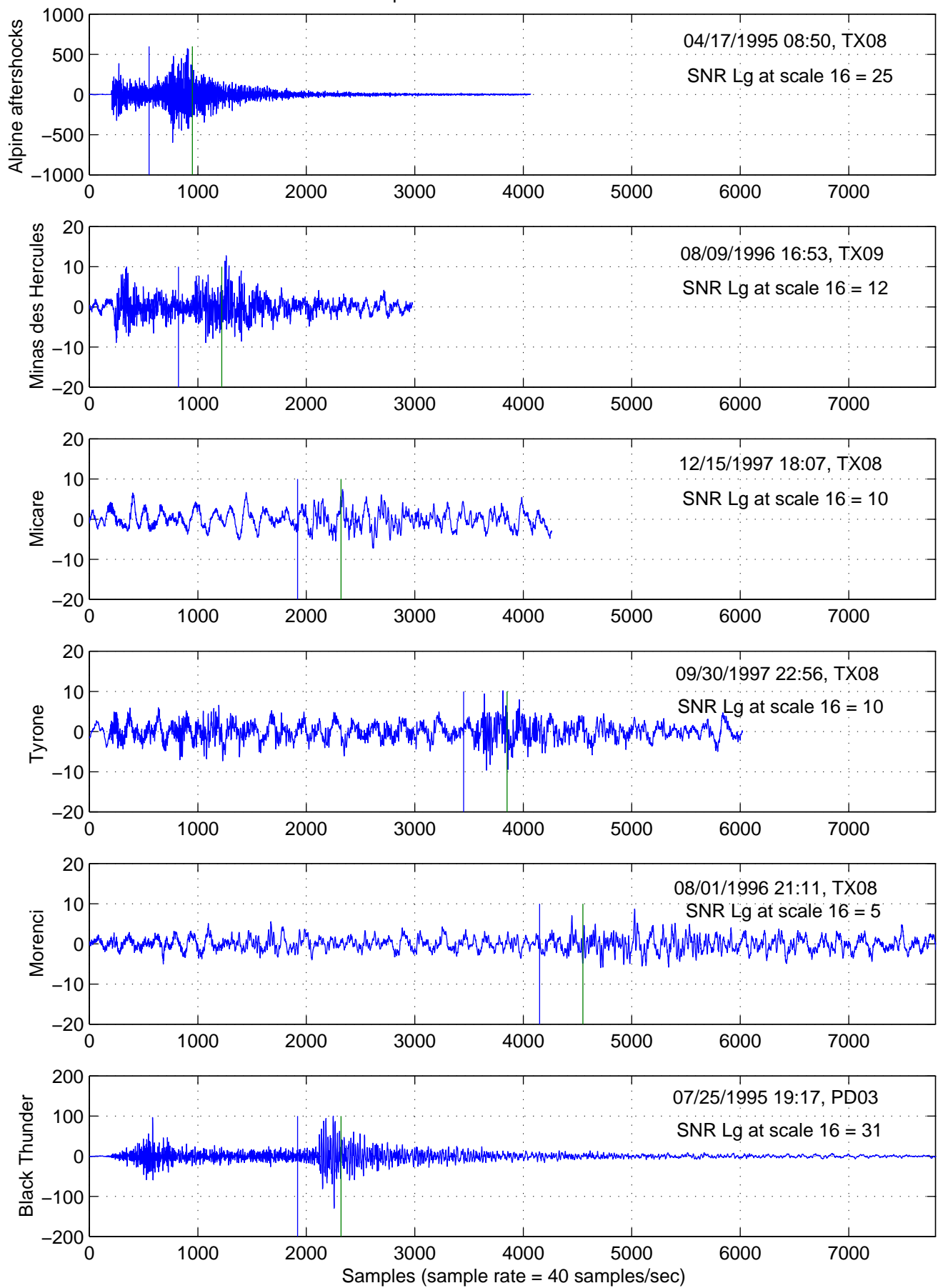

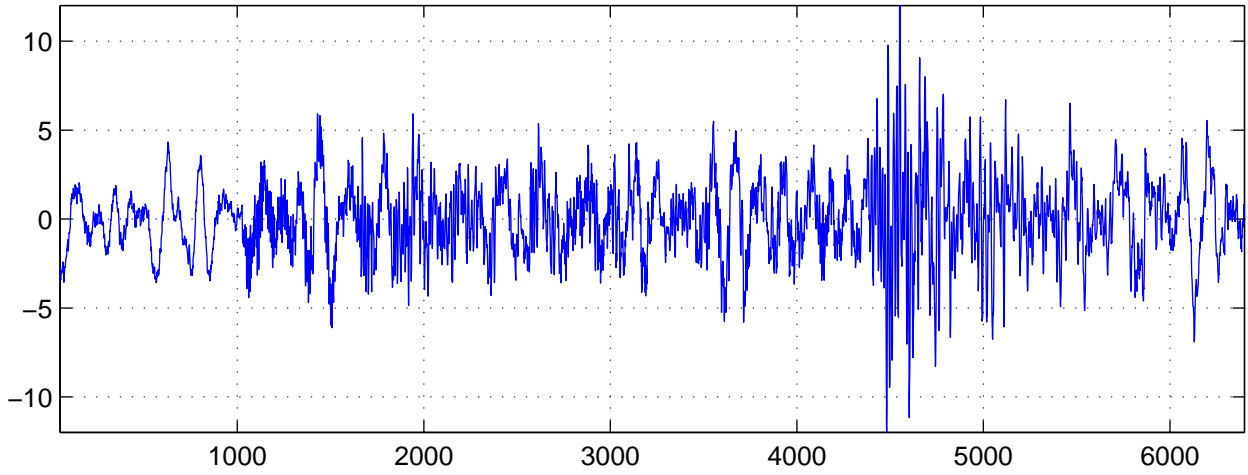
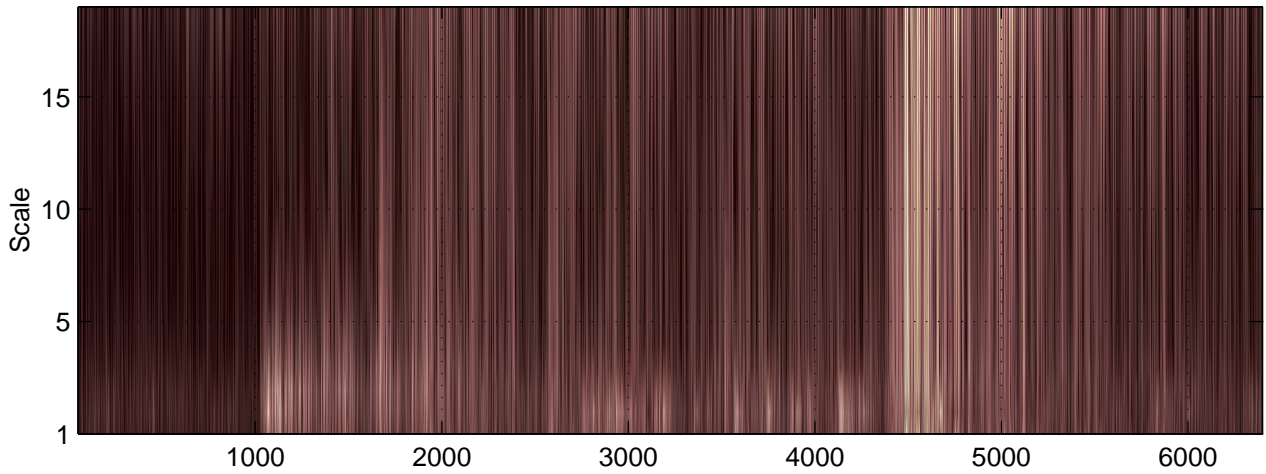
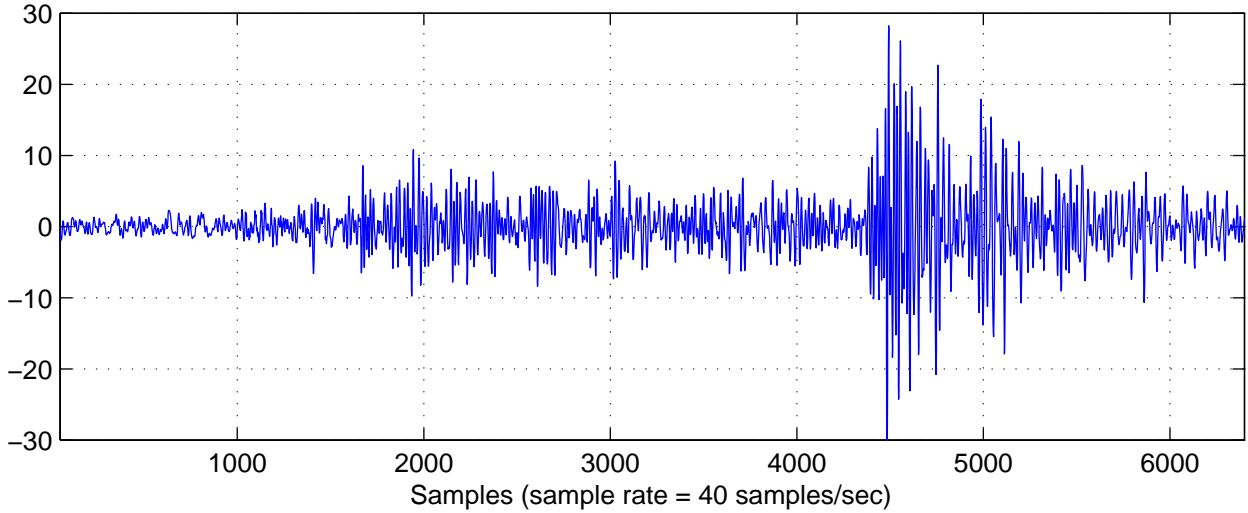

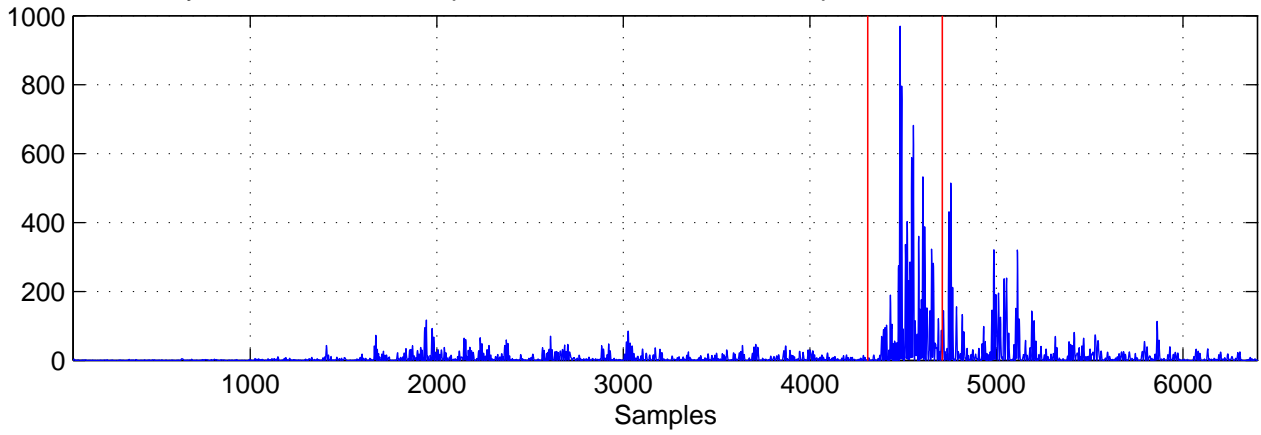

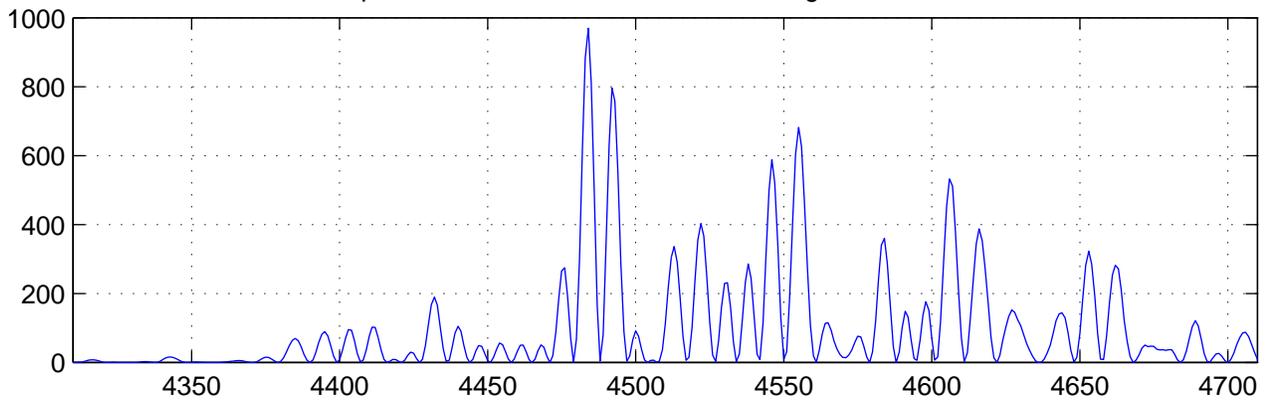

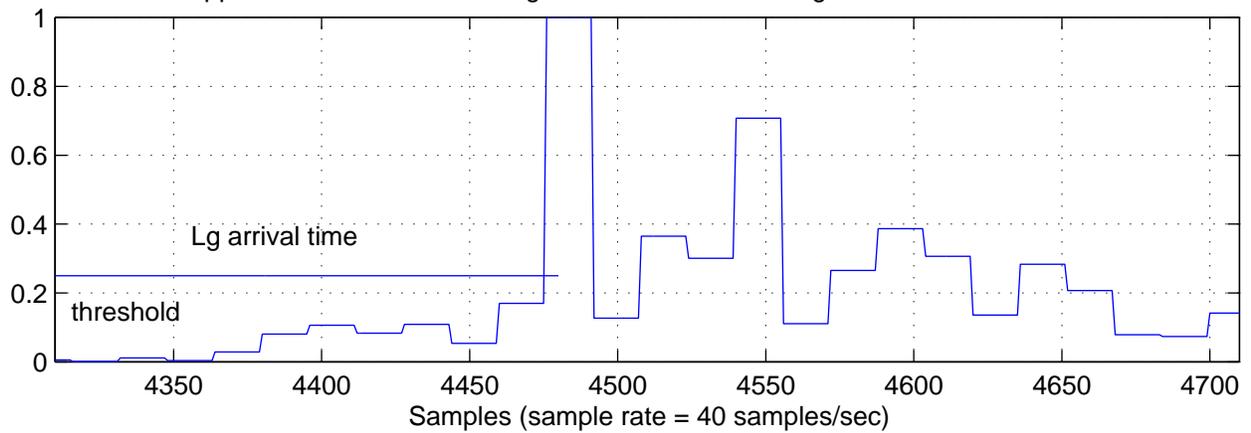

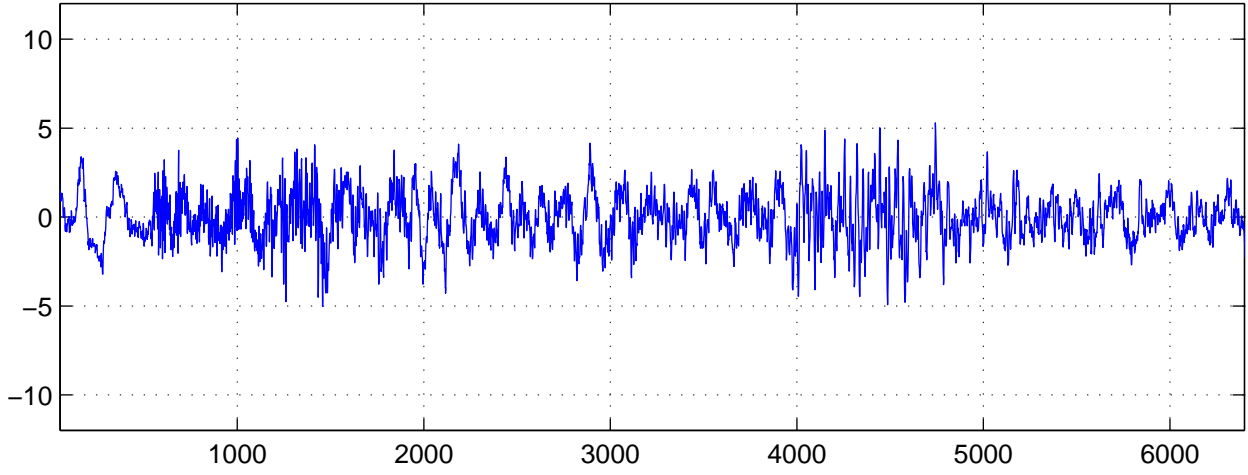
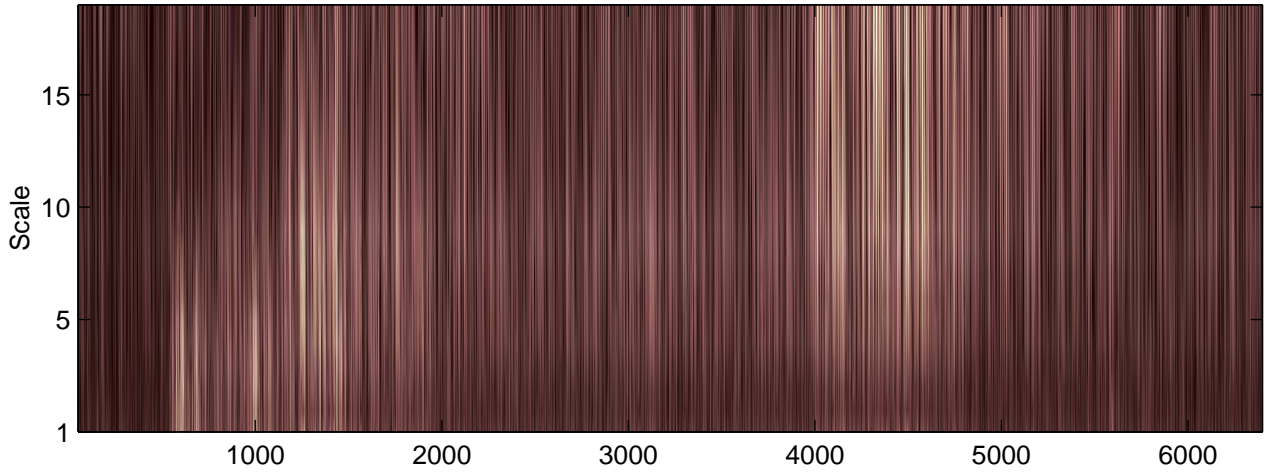
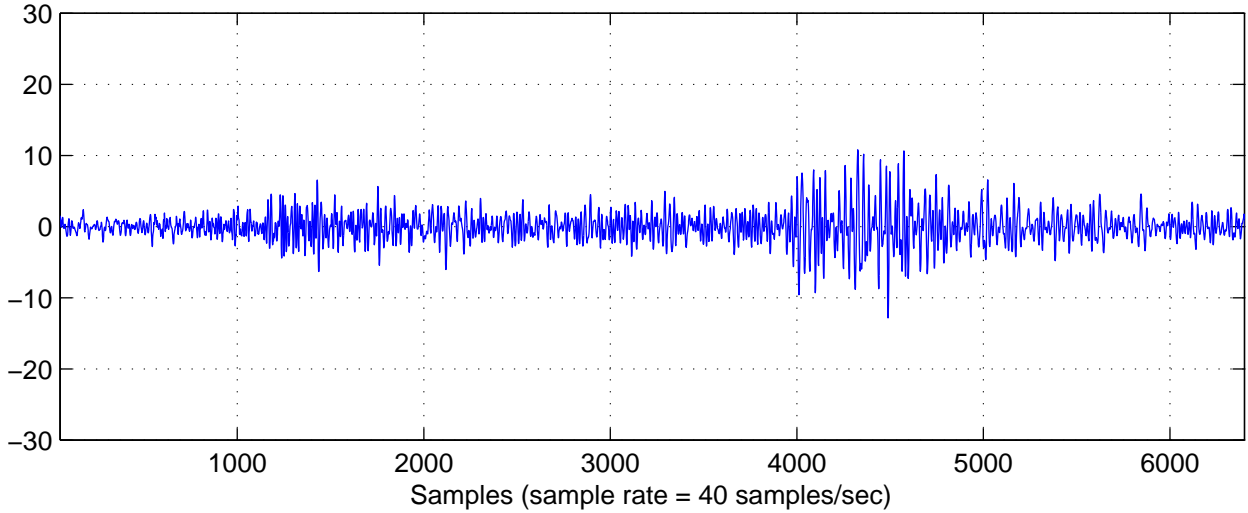

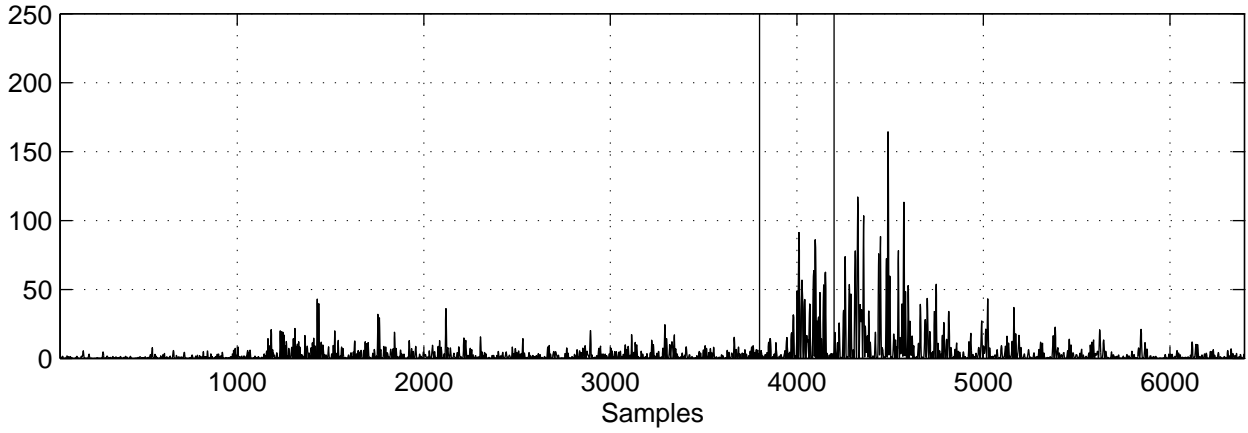

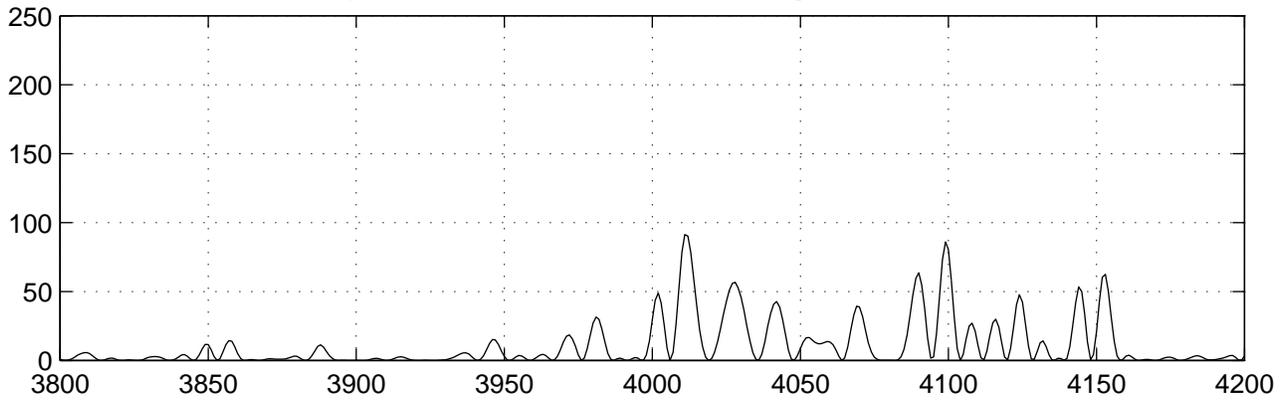

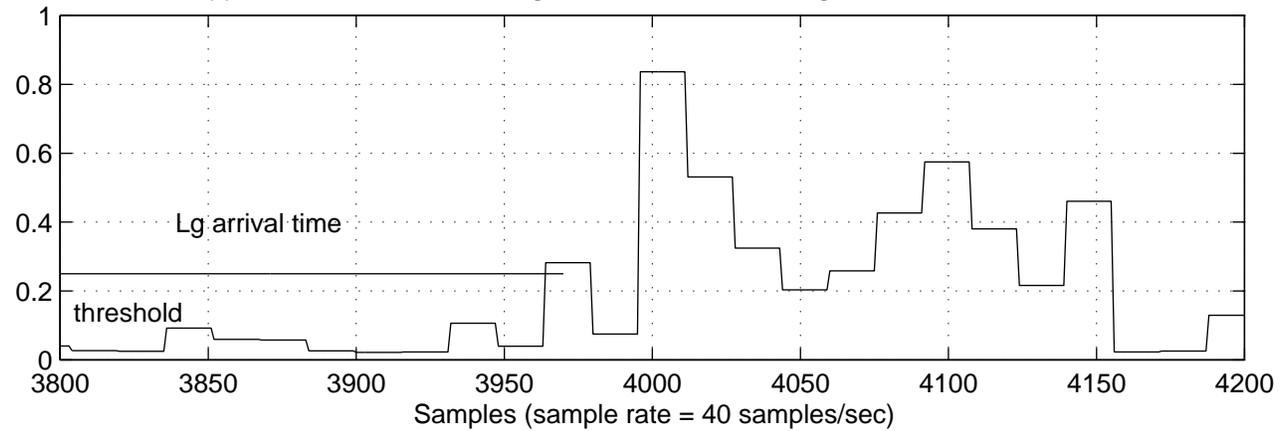

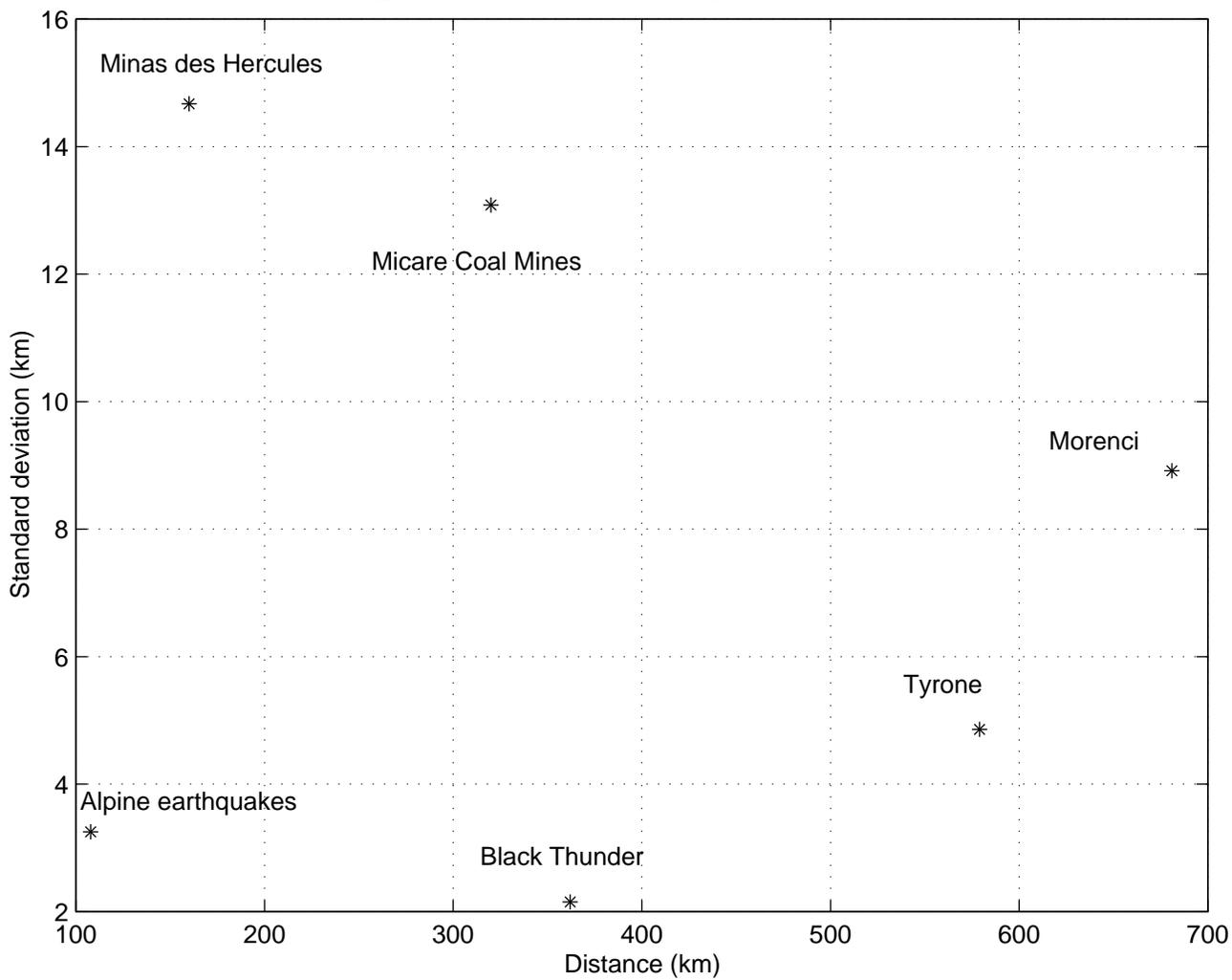

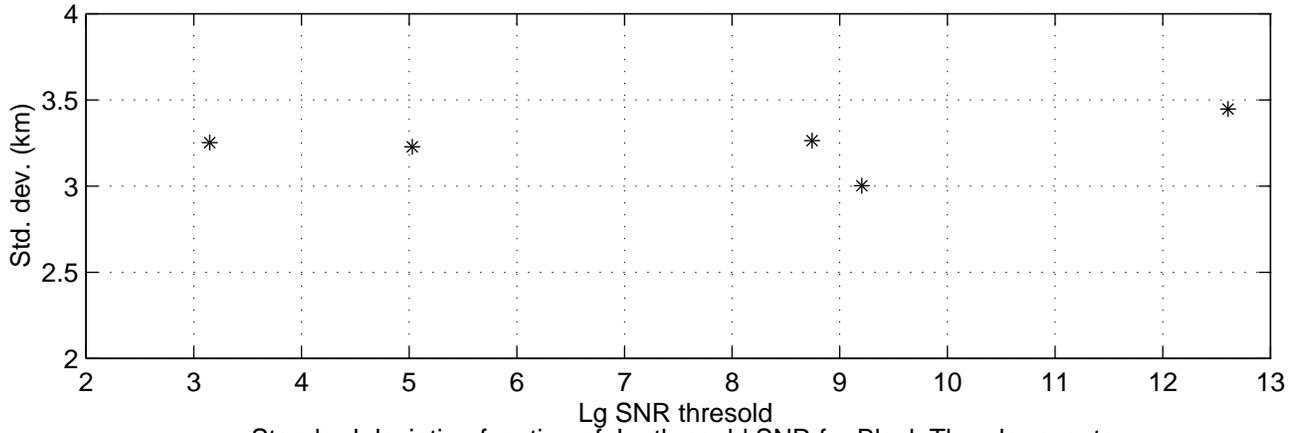
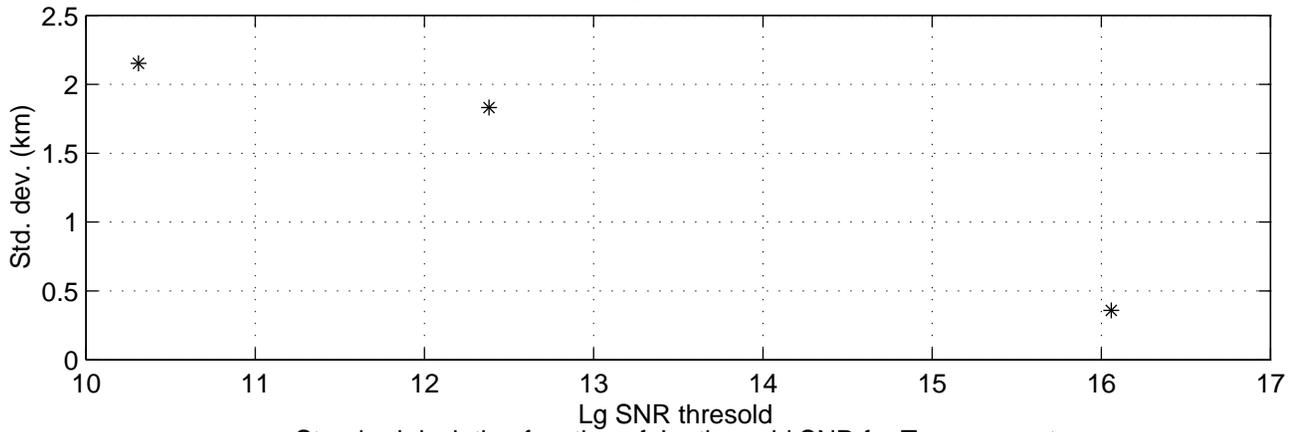
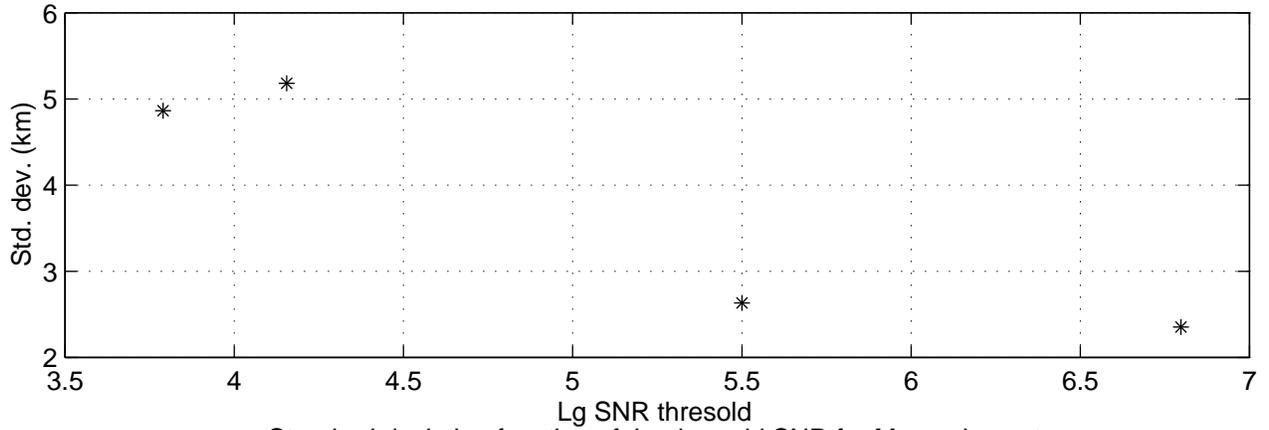
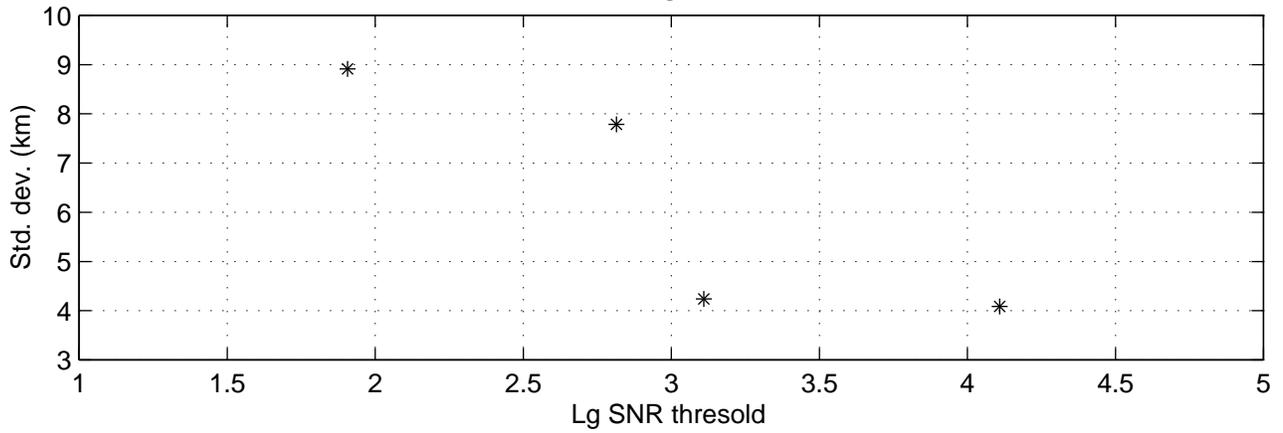

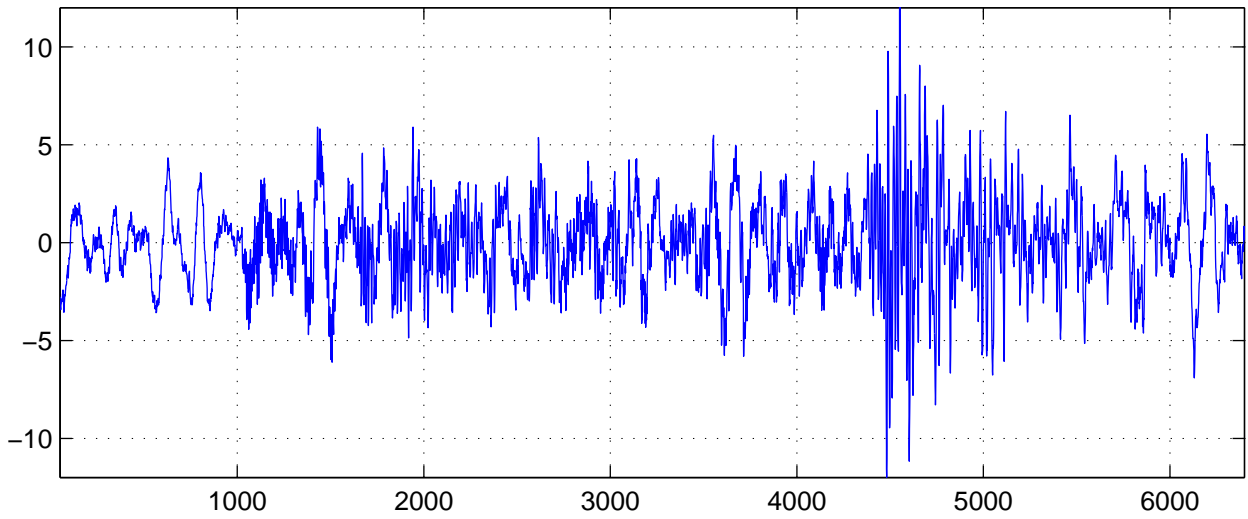
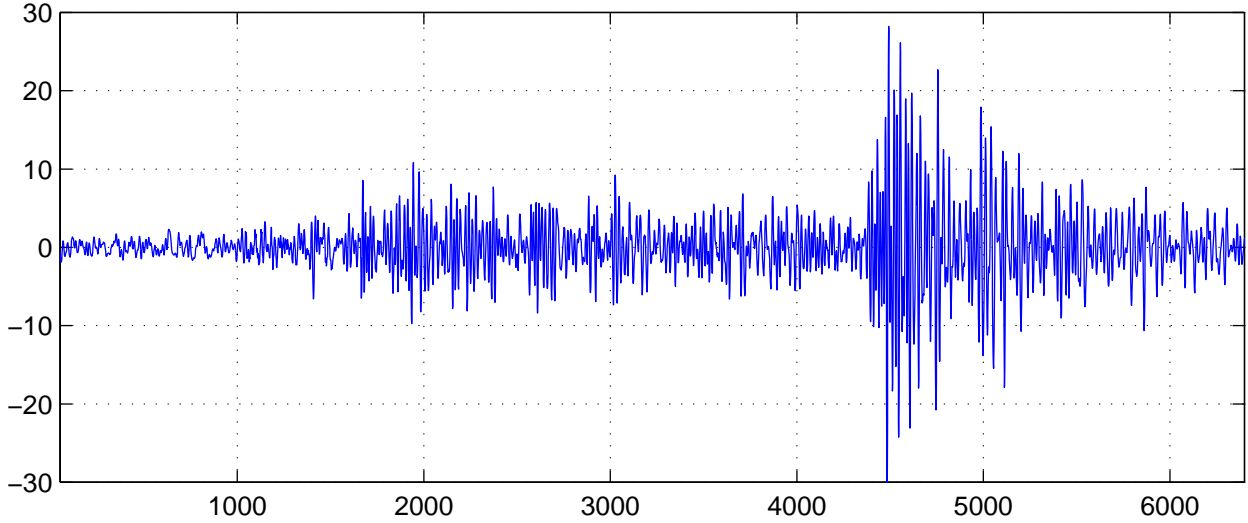
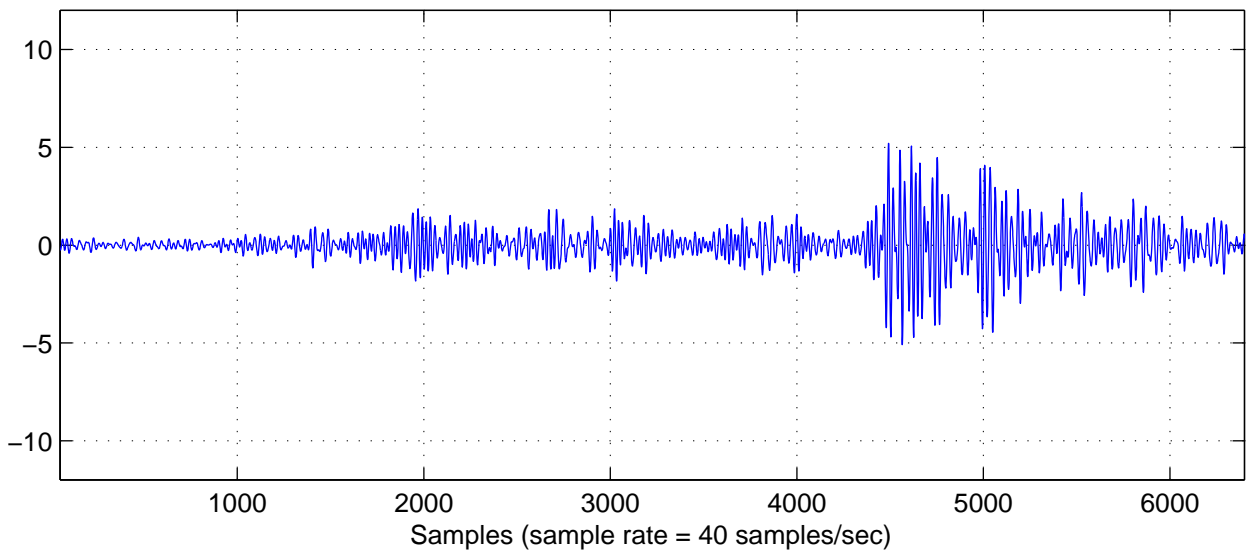

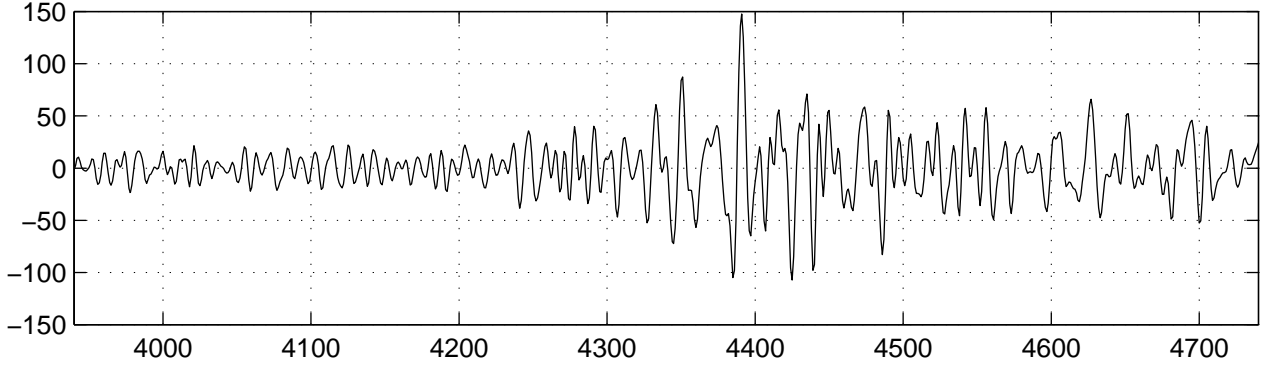
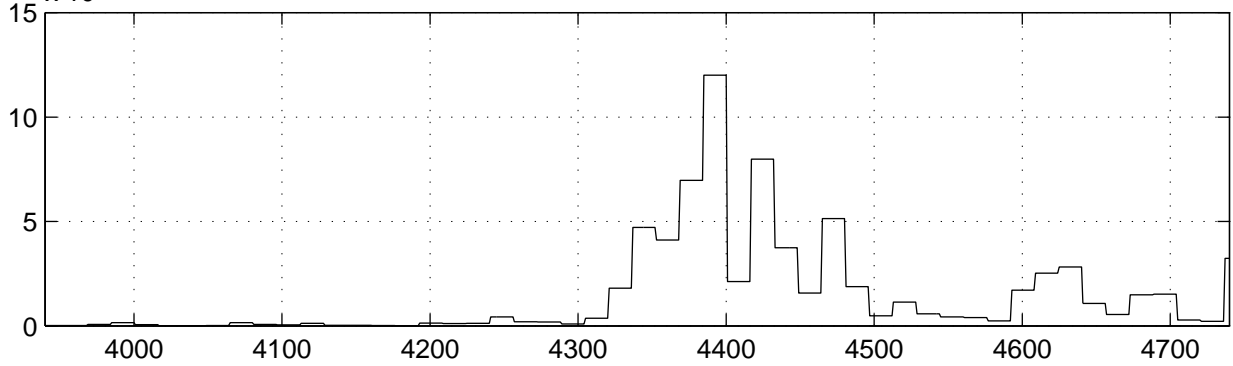
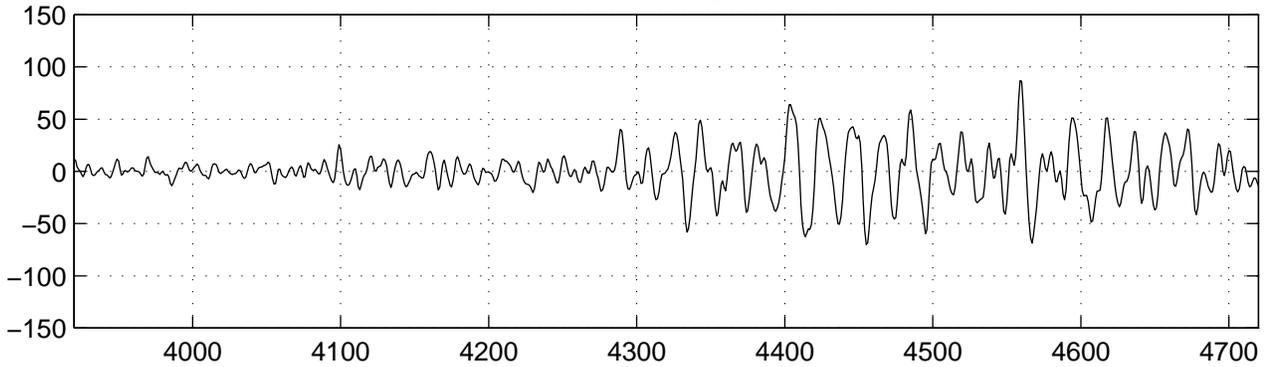
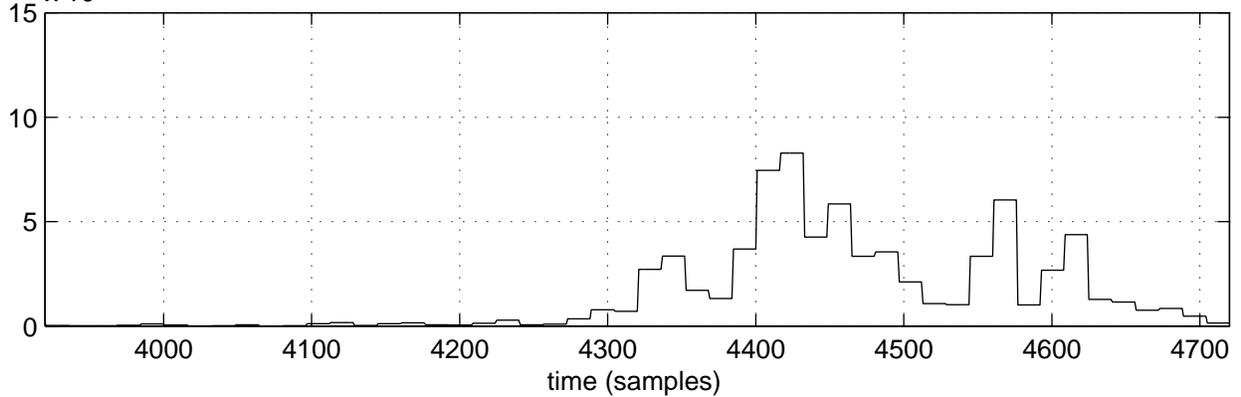